%% file: main.tex
\documentclass[%
 reprint,
superscriptaddress,
 longbibliography,
 amsmath,amssymb,
 aps,
 prl,
]{revtex4-2}

\usepackage[pagewise]{lineno}

\usepackage{graphicx}
\usepackage{dcolumn}
\usepackage{bm}
\usepackage[dvipsnames]{xcolor}
\usepackage[colorlinks,allcolors=BlueViolet]{hyperref}%

\begin{document}

\preprint{APS/123-QED}

\title{Single-dopant band bending fluctuations in \texorpdfstring{MoSe\textsubscript{2}\\}{MoSe2} measured with electrostatic force microscopy}

\author{Megan Cowie}
\affiliation{Department of Physics, McGill University, Montr\`eal, Qu\`ebec, Canada}
\author{Rikke Plougmann}
\affiliation{Department of Physics, Technical University of Denmark, Lyngby, Denmark}
\author{Zeno Schumacher}
\affiliation{Department of Physics, Institute of Quantum Electronics, ETH Zurich, 8093 Zürich, Switzerland}
\author{Peter Gr\"{u}tter}
\affiliation{Department of Physics, McGill University, Montr\`eal, Qu\`ebec, Canada}

\date{\today}

\begin{abstract}
In this work, we experimentally demonstrate two-state fluctuations in a metal-insulator-semiconductor (MIS) device formed out of a metallic atomic force microscopy tip, vacuum gap, and multilayer MoSe\textsubscript{2} sample. We show that noise in this device is intrinsically bias-dependent due to the bias-dependent surface potential, and does not require that the frequency or magnitude of individual dopant fluctuations are themselves bias-dependent. Finally, we measure spatial nonhomogeneities in band bending (charge reorganization) timescales. 
\end{abstract}

\maketitle

Individual charge state fluctuations have been observed in a variety of electrically isolated systems such as adatoms\cite{Gross2009}, quantum dots\cite{Bennett2010, RoyGobeil2015}, and molecules\cite{RoyGobeil2019} on insulators. Understanding these systems is critical for the study of single-electron physics, and two-state systems are of particular relevance for emerging quantum information technology. In semiconducting devices, individual charge states such as dangling bonds\cite{Turek2020}, individual dopants\cite{Teichmann2008}, and defects\cite{Plumadore2020} are not electrically isolated from their environment, and it is necessary to understand their effects on the global electronic structure, in particular device efficiency and noise. In this work, we measure single dopant fluctuations which give rise to variations in the surface potential of a mesoscopic metal-insulator-semiconductor (MIS) capacitor device. The MIS device is composed of a metallic frequency-modulated atomic force microscopy (fm-AFM) tip, a vacuum gap, and a multilayer MoSe\textsubscript{2} semiconducting sample.


The energy of a classical capacitor ($U_{ts}$) is found by summing the energies of the charge distributions of each capacitor plate: 

\begin{equation}\label{eq:U_ts}
    U_{ts}=\frac{1}{2} \left(\int_t \rho_t(z)V_t(z) \partial \tau_t + \int_s \rho_s(z)V_s(z)\partial \tau_s\right)
\end{equation}

\noindent where the first term is the energy of the top electrode (in this case, the fm-AFM tip) and the second term is the energy of the bottom electrode (MoSe\textsubscript{2} sample). $\rho_{t,s}(z)$ is the volume charge density, $\partial\tau_{t,s}$ is an infinitesimal volume element, and $V_{t,s}(z)$ is the potential of the tip or sample. For a metal-insulator-metal capacitor, $\rho$ and $V$ are spatially invariant, and the expression simplifies to $U_{ts}=\frac{1}{2}Q_{ts}V_{ts}=\frac{1}{2}C_{ts}V_{ts}^2$. The total charge ($Q_{ts}$) and potential difference ($V_{ts}$) are linear with applied bias and the capacitance ($C_{ts}$) is solely geometric. Deviations from a purely metallic system, such as upon introduction of static charges\cite{Kantorovich2000}, sample or tip polarizability\cite{Burke2009,Zhang2015}, and surface or interface dipoles\cite{Schmeits1996,Leon2016} can cause $\rho$ and $V$ to vary spatially and have a non-linear bias dependence. This is the case for semiconducting samples, where the surface potential $V_S$ that is established by the charge $Q_S$ inside a semiconductor is bias-dependent and spatially non-uniform (thus the potential that arises due to band bending.) In this work, $V_S$ has been calculated by numerically solving the following equation\cite{Hudlet1995}:

\begin{equation}\label{eq:Continuity}
    V_{S} =  V_{bias}+V_{\Phi}-\frac{Q_S}{C_I}
\end{equation}

\noindent where $Q_S=Q_S(V_S)$ and $C_I$ is the (geometric) capacitance per unit area, found by solving the Poisson Equation (see Section I of the Supplemental Material\cite{Supplemental} for this derivation). $V_{bias}$ is the applied bias between the tip and sample, and $V_\Phi$ is the difference in the tip and sample Fermi levels ($V_\Phi = E_{f,t}-E_{f,s}$). The total charge density is composed of thermal carriers, ionized dopants, and ``effective dopants", which are generically any ionizable states such as surface states, interface traps\cite{Kirton2006}, point vacancies\cite{Komsa2012}, contaminants, adatoms, interstitial atoms\cite{Seo2021}, etc. This incorporation of ionized ``effective dopant" states into the total charge density means that even an undoped sample can behave as though it were doped if other contributing effective dopant densities are appreciable. The occupation of dopant states can vary due to charge transfer from other sample locations, such as the substrate or insterstitial contamination layers\cite{Pollmann2020}. The MIS force ($F_{ts}$) per unit area ($a_{tip}$), derived in Hudlet \textit{et al} 1995\cite{Hudlet1995}, is:

\begin{equation}\label{eq:Force_MIS}
    \frac{F_{ts}}{a_{tip}} = \frac{Q_S^2}{2\epsilon}
\end{equation}

\begin{figure}[hb]
    \includegraphics[width=0.94\linewidth]{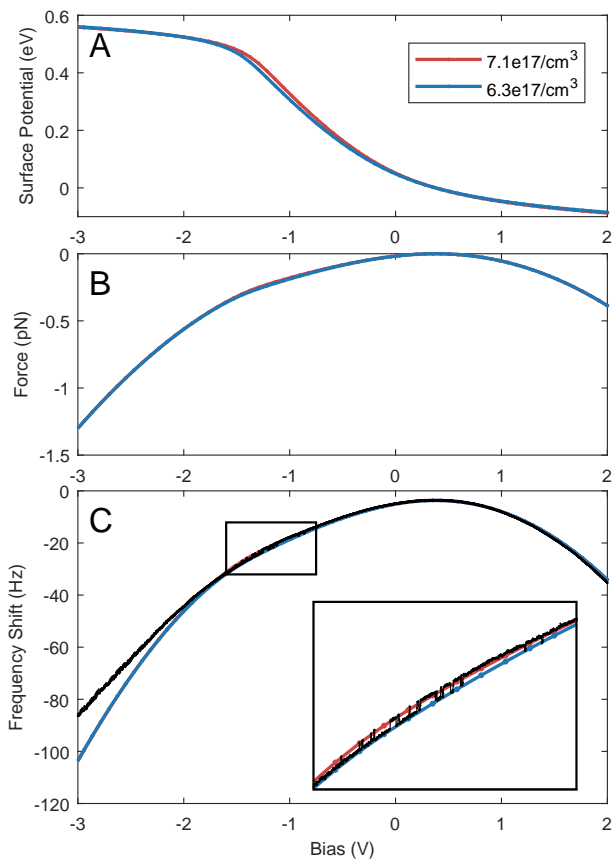}
    \caption{a) Modelled bias dependence of the MIS surface potential (Equation~\ref{eq:Continuity}) and b) force (Equation~\ref{eq:Force_MIS}) for two ionized acceptor concentrations:  6.3$\times$10\textsuperscript{17}/cm\textsuperscript{3} (blue) and  7.1$\times$10\textsuperscript{17}/cm\textsuperscript{3} (red). c) 
   Modelled fm-AFM frequency shift (blue and red, Equation~\ref{eq:FreqShiftDissipation}a) and data (black) on multilayer MoSe\textsubscript{2}. Inset: A zoom-in of the frequency shift for which fluctuations in the data (black) were observed. The experimental and model parameters for (a-c) are given in the main text. The sweep was acquired over approximately 10~seconds, and the bandwidth of the phase-locked-loop was 305~Hz.}  
    \label{fig:BiasSweeps}
\end{figure}

\noindent Figures~\ref{fig:BiasSweeps}a,b show the bias dependence of $V_S$ and $F_{ts}$ for two different ionized acceptor concentrations, where each data point is a numerical solution to Equations~\ref{eq:Continuity} and \ref{eq:Force_MIS}. The non-linear surface potential $V_S$ leads to a nonparabolicity in the force $F_{ts}$. 

In fm-AFM, a tip mounted on a cantilever (spring constant $k$, Q-factor $Q$) oscillates sinusoidally above a sample surface ($z=Asin\left(\omega t\right)$). Tip-sample force ($F_{ts}$) contributions which are in-phase with the cantilever motion lead to shifts ($\Delta\omega$) in the cantilever resonant frequency ($\omega_o$) and out-of-phase force contributions lead to variations in the cantilever drive ($F_d$), or excitation, required to maintain constant oscillation amplitude $A$:

\begin{subequations}\label{eq:FreqShiftDissipation}
\begin{eqnarray}
    \Delta\omega=\omega-\omega_o = \frac{-\omega_o}{2 kA}\frac{\omega_o}{\pi}\int_{0}^{2\pi/\omega}F_{ts}(t)sin(\omega t)\partial t~~~~ \\
    F_d = \frac{kA}{Q}-\frac{\omega_o}{\pi}\int_{0}^{2\pi/\omega}F_{ts}(t)cos(\omega t)\partial t ~~~~
\end{eqnarray}
\end{subequations}

\noindent (see Section II of the Supplemental Material\cite{Supplemental} for a derivation of Equation~\ref{eq:FreqShiftDissipation}.) The nonparabolicity in the force $F_{ts}$ which arises due to the nonlinearity in surface potential $V_S$ leads to a nonparabolic fm-AFM frequency shift above MoSe\textsubscript{2} (Figure~\ref{fig:BiasSweeps}c). Recently, similar nonparabolicities have been reported in other systems measured with fm-AFM, including dangling bonds on Si(111)\cite{Turek2020} and pentacene on KBr\cite{Schumacher2021}. Reference measurements on SiO\textsubscript{2}, in contrast, show a parabolic frequency shift as a function of bias. (See Section III of the Supplemental Material\cite{Supplemental} for these measurements.) 

The inset of Figure~\ref{fig:BiasSweeps}c shows a zoom-in of the kink in the frequency shift parabola. At these biases, the measured frequency shift (black) fluctuates between two states. This is due to individual dopant fluctuations which cause variations in $Q_S$. Notably, what is being measured here is the change in the global electrostatic environment (band bending, $V_S$) due to this single dopant fluctuation, and not a localized changing Coulomb interaction due to the fluctuating occupation of an isolated charge state. The former leads to the changes in nonparabolicity observed in this work, whereas the latter leads to parabola shifts, such as those shown in \cite{Gross2009}. This is supported by modelling: The two model fits shown differ only in their acceptor concentration, which are $6.3\mathrm{e}17/cm^3$ (blue) and $7.1\mathrm{e}17/cm^3$ (red). This is a dopant concentration difference of $0.8\mathrm{e}17/cm^3$, which corresponds to approximately one charge within the estimated $2.8\mathrm{e}\text{-}18~cm^3$ tip probing volume, assuming a probe volume equal to the effective tip area ($\pi\times10^2~nm^2$) times the sample thickness ($9~nm$). This fluctuation effect does not manifest homogenously over the sample surface; rather, it is only present at certain locations, and vanishes when the tip is moved slightly ($50~nm$). 

\begin{figure*}[ht]
    \includegraphics[width=\textwidth]{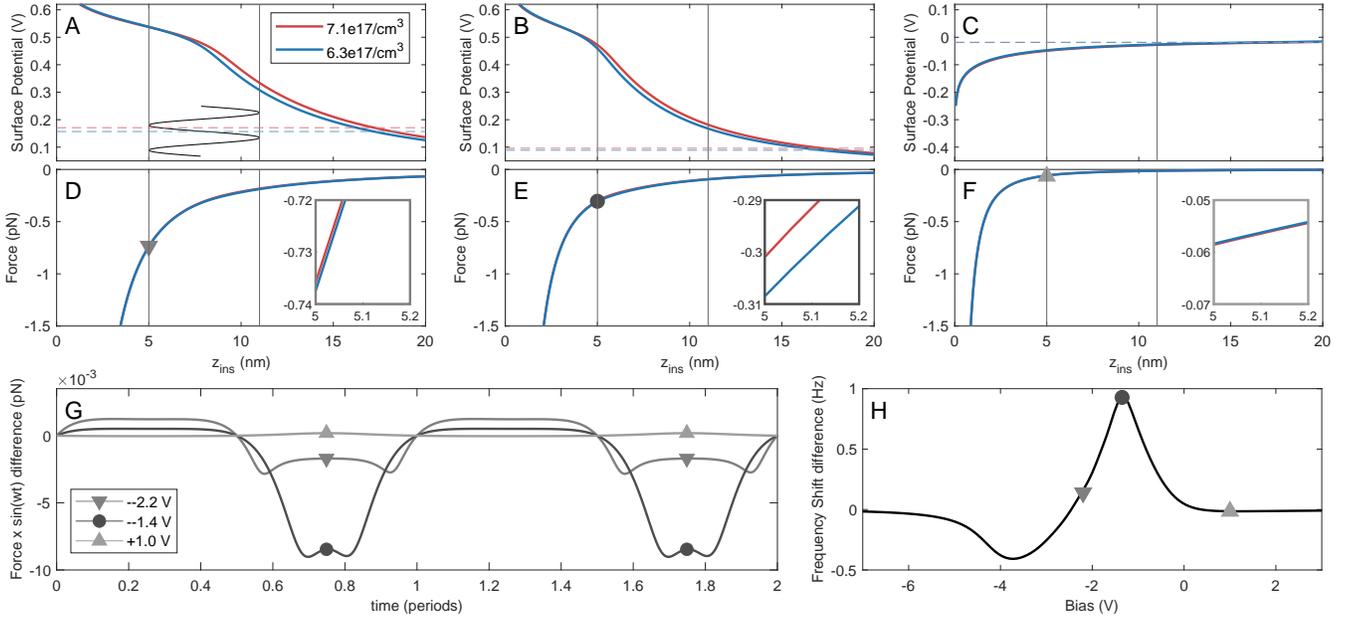}
    \caption{a-f) The surface potential and force as a function of tip-sample separation $z_{ins}$ at $-2.2V$ (a,d), $-1.4V$ (b,e), and $1.0V$ (c,f). Two different acceptor concentrations are shown: 6.3$\times$10\textsuperscript{17}/cm\textsuperscript{3} (blue) and  7.1$\times$10\textsuperscript{17}/cm\textsuperscript{3} (red). Vertical black lines indicate the region of the curves over which the cantilever oscillates with an amplitude of $6~nm$ and closest $z_{ins}$ of $5~nm$ (a demonstrative sinusoid is shown in (a)). In (a-c), horizontal lines are drawn to highlight the surface potential at the closest tip-sample separation for each acceptor concentration (in a and c they are overlapping). In (d-f), an inset shows the closest $z_{ins}$ force at the position indicated by the marker. g) The difference between the two acceptor concentrations of the integrand of Equation~\ref{eq:FreqShiftDissipation}a ($F_{ts}\times\sin(\omega t)$) over two oscillation cycles at $-2.2~V$, $-1.4~V$, and $1.0~V$. h) The frequency shift difference between the two acceptor concentrations, with representative biases indicated with markers.}
    \label{fig:DifferencePeak}
\end{figure*}

The measured frequency shift fluctuation is maximized at biases corresponding to the kink in the surface potential ($\sim -1.4~V$). This is demonstrated in Figure~\ref{fig:DifferencePeak}, which shows the surface potential and force as a function of tip-sample separation ($z_{ins}$) for two acceptor concentrations at $-2.2~V$ (a,d), $-1.4~V$ (b,e), and $1.0~V$ (c,f). 
Equation~\ref{eq:FreqShiftDissipation}a indicates that the closest $z_{ins}$ force gives the largest contribution to the fm-AFM frequency shift, due to the multiplication of $F_{ts}$ by $\sin(\omega t)$. The difference in this integrand for the two acceptor concentrations is shown over two oscillation cycles in Figure~\ref{fig:DifferencePeak}g. Integrating Figure~\ref{fig:DifferencePeak}g according to Equation~\ref{eq:FreqShiftDissipation}a gives Figure~\ref{fig:DifferencePeak}h. Figure~\ref{fig:DifferencePeak}h is effectively a ``noise sensitivity function" for these two-state fluctuations.

Highlighted by horizontal lines in Figure~\ref{fig:DifferencePeak}a-c and markers with insets in d-e are the surface potential and force at the closest tip-sample separation. At $-1.4~V$, the two acceptor concentrations have a comparatively large difference in surface potential and force at the closest $z_{ins}$, and consequently the difference of $F_{ts}\times \sin(\omega t)$ (the integrand of Equation~\ref{eq:FreqShiftDissipation}a) is largest.  In comparison, at $-2.2~V$ there is a large difference in surface potential at the top of the oscillation cycle, but since it is not maximally amplified by $\sin(\omega t)$, the resulting frequency shift difference is small.


This model, therefore, shows that the bias dependence of the frequency shift fluctuation amplitude (Figure~\ref{fig:BiasSweeps}c) is \textit{not} due to a bias dependence of the dopant state (which has nowhere been incorporated into this model). Rather, it is due to an amplification from the intrinsic nonlinear surface potential: At biases corresponding to the parabola kink, where the surface potential has the largest nonlinearity, the difference between the ionized state force and the unionized state force is maximized at the closest tip-sample position, leading to the largest difference in frequency shift.

\begin{figure}[b]
    \centering
    \includegraphics[width=\linewidth]{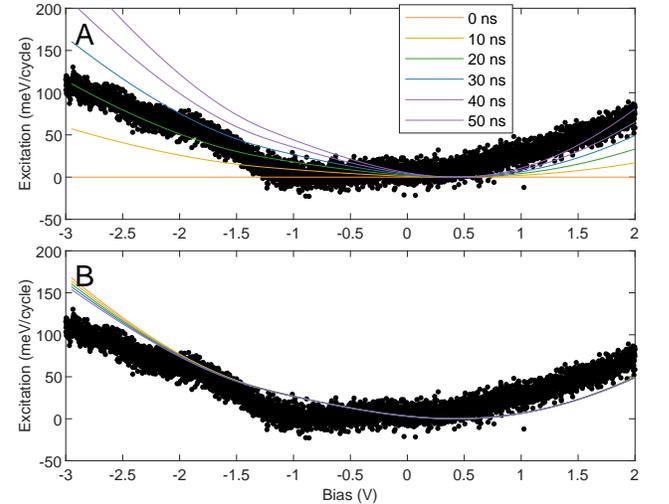}
    \caption{Modeled excitation signal as a function of bias with a) constant Q-factor ($18000$); and b) constant lag ($30~ns$). The other parameters match those defined in the main text. Experimental results taken simultaneously with those shown in Figure~\ref{fig:BiasSweeps} are shown in black.}
    \label{fig:FitParameters_Lag}
\end{figure}

Figure~\ref{fig:FitParameters_Lag} shows the excitation signal measured simultaneously with the frequency shift data shown in Figure~\ref{fig:BiasSweeps}c. The non-constant excitation signifies an out-of-phase tip-sample force $F_{ts}$, according to Equation~\ref{eq:FreqShiftDissipation}b. This out-of-phase component arises from charge reorganization inside the sample, or in other words, it is an equilibration timescale for the surface potential. This lag time is modeled by incorporating a phase offset between the closest tip sample position and the maximum tip-sample force. Modeled lag times ranging from $0$ to $50~ns$ are shown in Figure~\ref{fig:FitParameters_Lag}a, and the best estimate of this band bending timescale is $30~ns$. This timescale is effectively a resistance-capacitance (RC) time constant for the tip-vacuum-sample MIS system. An order-of-magnitude estimate of the time constant $\tau=RC$, assuming a simple parallel-plate capacitance $C=\frac{\epsilon \epsilon_o a}{z_{ins}}$ and resistivity $\rho = \frac{Ra}{z_{ins}}$ (where $\epsilon$ and $\epsilon_o$ are the relative and free permittivity, and $a$ and $z_{ins}$ are the plate area and separation distance), agrees with this model: Using the best-fit  relative permittivity from this work ($\epsilon=5.9$) and a resistivity of $500~\Omega m$ (which is within the wide reported resistivity range of $0.1-1000~\Omega m$ for MoSe\textsubscript{2}\cite{Zhu2012}),  gives $\tau = 26~ns$, which is consistent with the measured $30~ns$. The model appears to be simplistic as the fit misses many details of the data. This could be due to a bias-dependent lag time, which has not been accounted for here. Nonetheless this explanation can be useful for an order-of-magnitude estimate of the band-bending timescale. Note that when the lag is zero, the excitation signal is flat, and this is independent of Q. Furthermore, Figure~\ref{fig:FitParameters_Lag}b demonstrates that this excitation signal does not arise from variations in Q as a function of bias, since large variations in Q at a constant lag time ($30~ns$) do not reproduce the measured phenomenon. This confirms that the measured excitation signal is not due to `dissipation' (i.e. a non-conservative force), but rather a time-delayed conservative force. In comparison, reference measurements on SiO\textsubscript{2}, shown in Section III of the Supplemental Material\cite{Supplemental} show constant excitation as a function of bias.

\begin{figure}[b]
    \centering
    \includegraphics[width=\linewidth]{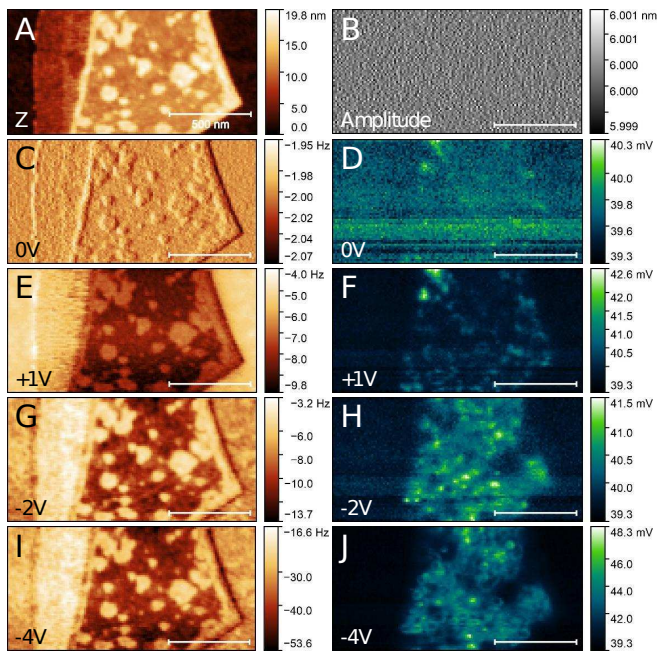}
    \caption{Multipass fm-AFM image (i.e. constant frequency shift image) of a region of an exfoliated MoSe\textsubscript{2} island. a-d) Topography (a), oscillation amplitude (b), frequency shift (c), and excitation (d) channels recorded during the first pass, at $0~V$ and $-2~Hz$ setpoint. The topography shown in (a) is re-traced for each subsequent pass. The frequency shift (left) and excitation (right) channels are shown for the subsequent passes, which are at $+1~V$ (e-f), $-2~V$ (g-h), and $-4~V$ (i-j). The horizontal scale bars correspond to $500~nm$.}
    \label{fig:Multipass}
\end{figure}

The excitation channel can be measured spatially to map variations in band bending timescales over a sample surface. This is demonstrated in Figure~\ref{fig:Multipass}, which shows multipass frequency shift and excitation images at varying bias. The highest layer of the island, susceptible geometrically to the largest surface potential since it is not so spatially limited in the z direction, exhibits the most appreciable bias dependence in both frequency shift and excitation. At $-2~V$, bright spots appear in the excitation channel, indicating a spatial nonhomogeneity in the charge equilibration time. The charge regorganization timescales at these sites are largest near $-1.4~V$ because this bias corresponds to the largest change in surface potential over every oscillation cycle (as demonstrated in Figure~\ref{fig:DifferencePeak}). At larger negative voltages, rings appear in the excitation channel. These rings further indicate the spatial and bias dependence of the charge reorganization time: When the tip is laterally offset from a long reorganization timescale site, there is an additional potential drop. These rings, visible in Figure~\ref{fig:Multipass}j at $-4~V$, therefore correspond to the same phenomenon shown in Figure~\ref{fig:Multipass}h at $-2~V$.

We note that this observation of bias-dependent excitation is not a piezoacoustic excitation system transfer function artefact such as described in \cite{Labuda2011}, since by changing the position slightly on the sample, the frequency shift and excitation vary independently (this is shown in the Supplemental Material\cite{Supplemental}). This excitation signal is also not due to tunneling charge transfer between the sample and tip, since the effect was observed at tip-sample separations $>5~nm$ (also see Supplemental Material\cite{Supplemental}).


\textit{Model Description --} The surface potential, force, frequency shift, and excitation were modelled according to Equations~\ref{eq:Continuity}, \ref{eq:Force_MIS}, and \ref{eq:FreqShiftDissipation}. Over every oscillation cycle, the tip-sample separation changes, the potential drop across the vacuum gap changes, and the potential drop inside the semiconductor changes, resulting in an time-dependent surface potential and force. This time-dependent force is integrated to determine the frequency shift and excitation, where the excitation signal is only non-zero when there is a band-bending timescale, and consequently a phase offset between the closest tip-sample position and the maximum tip-sample force. See Section IV of the Supplemental Material\cite{Supplemental} for more detailed explanation of this process, and the Supplemental Material\cite{SupplementalVideo} for a video description. Each model point shown in Figures~\ref{fig:BiasSweeps}a-c and \ref{fig:FitParameters_Lag} is a numerical solution to Equations~\ref{eq:Continuity}, \ref{eq:Force_MIS}, and \ref{eq:FreqShiftDissipation} at a different bias.

\textit{Model Parameters --} Equations~\ref{eq:Continuity}, \ref{eq:Force_MIS}, and \ref{eq:FreqShiftDissipation} involve 14 experimental parameters. The impact each has on the overall parabola shape is distinct: The band gap, acceptor concentration, permittivity, and temperature introduce nonparabolicity; the tip work function and sample electron affinity introduce lateral shifts and leave the shape unchanged; and the tip-sample separation, tip radius, oscillation amplitude and frequency, spring constant, and Q-factor are all fm-AFM scaling factors that are multiplicative prefactors which do not change the overall shape. See Section V of the Supplemental Material\cite{Supplemental} for a further explanation of the sensitivity of each of these parameters and their impact on the shape. Optimal fit values were found by exploring a large parameter space (\textgreater~120,000 curves) and finding the fit that minimizes residuals. The few-layer MoSe\textsubscript{2} band gap is $1.55~eV$\cite{Labuda2011,Miyahara2015} (best fit $1.5~eV$), electron affinity $3.5~eV$\cite{Kantorovich2004} (no fit against this parameter), and relative (dielectric) permittivity $5.7$\cite{Holscher2001} (best fit $5.9$). A rigorous interpretation of permittivity is challenging because there is likely water in between the thin flaked MoSe\textsubscript{2} sample and the SiO\textsubscript{2} substrate. However, taking this fitted permittivity as an ``effective permittivity" of the net system does not affect the overall interpretation. The effective electron and hole masses were assumed to be $1.0$, and the temperature $300~K$. The spring constant of $42~N/m$ estimated by the tip manufacturer agreed with the best fit. The Q-factor $18000$ and resonant frequency $330~kHz$ were measured experimentally by performing a frequency sweep. The oscillation amplitude setpoint, maintained with a phase-locked loop and a feedback circuit, was $6~nm$. The work function of Si/SiO\textsubscript{2} is $4.4~eV$, and given that the substrate parabola peak (i.e. $V_{\Phi}$, see Supplemental Material Section III\cite{Supplemental}) is $-250~mV$, and $V_{\Phi}$ is the difference in the tip and sample work functions ($V_{\Phi}=\Phi_{tip} - \Phi_{sample}$), $\Phi_{tip}$ is $4.15~eV$ (best fit $4.1~eV$). The effective tip radius was taken to be $10~nm$, which is consistent with the tip manufacturer's estimate of an effective radius ``better than $25~nm$". The optimal closest tip-sample separation $z_{ins}$ was $5.2~nm$. The position of the experimental parabola kink at negative biases indicates that the effective dopants are p-type. The dopant concentration, the final fitting parameter, is optimized at $6.3 \mathrm{e}17/cm^3$ (blue) and $7.1 \mathrm{e}17/cm^3$ (red). The deviation at large negative biases in Figures~\ref{fig:BiasSweeps}c-e could be due to the emergence of stray capacitances as the applied electric field increases.

\textit{Experimental Parameters --} The sample of MoSe\textsubscript{2} on SiO\textsubscript{2} measured in this work was prepared by all-dry viscoelastic stamping\cite{Castellanos2014}. The top layer of the micron-scale multilayer island of MoSe\textsubscript{2} is $9.0~\pm~0.4~nm$ above the silicon substrate (see \cite{Plougmann2021} for height measurement methodology.) All fm-AFM measurements were taken using Nanosensors platinum-iridium coated silicon tips (PPP-NCHPt) with $330~kHz$ resonant frequency, spring constant $42~N/m$, and Q factors approximately $18000$ in a JEOL JSPM-4500A UHV AFM with a Nanonis control system. All measurements were performed in ultra-high vacuum (base pressure $ < 3\times10^{-10}~mbar$) at room temperature. The sample was annealed at $120^{\circ}C$ for eight hours each time it was introduced into vacuum, and was grounded during all measurements.


In conclusion, we have experimentally demonstrated the direct relationship between a single fluctuating dopant state and its effect on the global band structure (band bending) in a mesoscopic MIS device. The bias dependence of these fluctuations does not depend on the bias dependence of the dopant state occupation, but is rather inherently due to the bias dependence of the surface potential. This has important ramifications for MIS-like device functionality and noise: It indicates that even in the absence of bias dependent dopant or defect states, device noise is bias dependent. The fluctuating two-state effect is demonstrated here for the well-characterized two-dimensional MoSe\textsubscript{2} system, but it has also been observed in pentacene (publication in preparation). This indicates that this is not a sample-specific phenomenon, but rather is relevant for a thorough understanding of noise in any semiconductor device. Additionally, we have demonstrated that band bending equilibration timescales may be measured using the fm-AFM excitation signal. Given that fm-AFM affords high spatial resolution, this approach may be used to direcly measure band bending timescales of different types of defects. 

We thank NSERC, FRQ-NT and CFI for funding, Alexander~Schluger, Kirk Bevan, and Hong Guo for stimulating discussions, and Philipp~Nagler for the preparation of the MoSe\textsubscript{2} sample.


\nocite{*}
\bibliography{mainbib}


\input{Supplemental}

\end{document}

%% file: Supplemental.tex
\newpage
\onecolumngrid
\pagenumbering{gobble}
\newpage

\begin{center}
\huge Supplementary Materials
\end{center}
\bigskip 

\section{I. Surface Potential and Force} \label{Sec:SurfacepotForce}
When a bias is applied to a semiconductor, its finite density of states means that there is a spatially-dependent electric field inside the semiconductor. This causes charge within the semiconductor to re-organize, and creates a spatially dependent potential that decays with distance (band bending). The amount of band bending, that is, the difference between the bulk potentials and the potential at the surface, is the surface potential, and is calculated by solving the Poisson equation: 

\begin{equation}\label{eq:Poisson1}
    \frac{\partial\vec{E}}{\partial z}=-\nabla^2 V = \frac{\rho}{\epsilon}
\end{equation}

\noindent where the total charge density in the semiconductor $\rho$ is generally taken to be due to the thermally excited carriers ($n(z)$, $p(z)$) and ionized dopants ($n_D^+$, $n_A^-$):

\begin{equation}\label{eq:Poisson2}
   \rho = e\left(n(z)-p(z)+n_D^+-n_A^-\right)
\end{equation}

\noindent Imperfect (real-world) samples with surface or defect states, such as edge states, dangling bond states, etc, can have additional carriers contributed by either filling ($p_s$) or depleting ($n_s$) defect states, just as dopants do. A robust definition of $\rho$ therefore includes these contributions:

\begin{equation}
\begin{aligned}
   \rho &= e\left(n(z)-p(z)+n_D^+-n_A^-+n_d^++n_a^-\right) \\
   &= e\left(n(z)-p(z)+n_{D,d}^+-n_{A,a}^-\right) \\
\end{aligned}
\end{equation}

\noindent such that the Poisson equation is: 

\begin{equation}
   \frac{\partial\vec{E}}{\partial z} = \frac{e}{\epsilon} \left(n(z)-p(z)+n_{D,d}^+-n_{A,a}^-\right)
\end{equation}

The charge inside the semiconductor is\cite{Hudlet1995}:
\begin{equation}
    Qs = sgn(u) \frac{\epsilon\epsilon_o k_BT }{eL_D}  \left[e^u-u-1+\frac{n_i^2}{N_A^2}\left(e^{-u}+u-1\right)\right]^{1/2}
\end{equation}

\noindent where $u = \frac{eV_S}{k_BT}$ and $L_D = \sqrt{\frac{\epsilon\epsilon_o k_BT}{2N_Ae^2}}$. The surface potential is found by solving the following continuity equation numerically\cite{Hudlet1995}: 

\begin{equation}\label{eq:Suppcont}
    0 = V_{bias}+V_{\Phi}-V_S-\frac{Q_S\times z_{ins}}{\epsilon_o}
\end{equation}

\noindent where $V_{\Phi} = E_{f,t}-E_{f,s}$ and $E_{f,t}$ and $E_{f,s}$ are the tip and sample Fermi levels, respectively. Finally, the net tip-sample force is found algebraically given the $V_S$ solution:

\begin{equation}\label{eq:SuppFts}
    F_{ts} = a_{tip}\times\frac{Q_S^2}{2\epsilon_o}
\end{equation}

\noindent where $a_{tip}=\pi r_{tip}^2$ is the effective tip area. This derivation is shown in greater detail in Hudlet \textit{et al} (1995)\cite{Hudlet1995}. (Note that the expressions have been modified slightly to account for the fact that this system is p-type, and accounting for the fact that the tip and sample have different Fermi levels.)

\newpage 
\section{II. Frequency Shift and Excitation} \label{Sec:FreqshiftExcitation}

The following section demonstrates that in an fm-AFM experiment, the frequency shift arises from the in-phase component of the tip-sample force, while variations in excitation arises from the out-of-phase component. (This derivation is also presented elsewhere\cite{Holscher2001,Kantorovich2004,Sader2005,Miyahara2015}, but is included here for clarity.) In the small amplitude limit, an fm-AFM cantilever behaves like a damped, driven harmonic oscillator with equation of motion:

\begin{equation}
 m\ddot{z} + \xi\dot{z} + kz = F_{drive} + F_{ts}(z,t)
\end{equation}

\noindent where $m$ is the cantilever mass, $\xi$ is the damping coefficient, $k$ is the spring constant,  $F_{drive}$ is the driving force, and $F_{ts}(z,t)$ is the tip-sample interaction force. Rewritten in terms friendlier to an experimentalist, given that the cantilever resonance frequency $\omega_o=\sqrt{\frac{k}{m}}$ and quality factor $Q=\frac{m\omega_o}{\xi}$, we have:

\begin{equation}
 k\ddot{z} + \frac{k\omega_o}{Q}\dot{z} + k\omega_o^2z = \omega_o^2(F_{drive} + F_{ts}(z,t))
\end{equation}

\noindent We can solve the equation of motion using the ansatz $z=A\sin(\omega t)$. The drive, which when optimized is 90 degrees out of phase with the position, is $F_{drive}=F_d \cos(\omega t)$), such that: 

\begin{equation}\label{eq:eom_substituted1}
    F_{ts}(t)=\left[\frac{kA}{\omega_o^2}(\omega_o^2-\omega^2)\right]\sin(\omega t)+\left[\frac{kA}{\omega_o^2}\frac{\omega\omega_o}{Q}-F_d\right]\cos(\omega t)\\
\end{equation}

\noindent The total force $F_{ts}$ can be generically expressed as a Fourier series with the orthonormal basis of $\sin(t)$ and $\cos(t)$. Taking only the terms at frequency $\omega$, we find that the force can be expressed as:

\begin{equation}
    F_{ts}(t)=F_{in} \sin(\omega t) + F_{out} \cos(\omega t)
\end{equation}

\noindent where the amplitudes of the in-phase and out-of-phase force contributions are the Fourier coefficients:

\begin{subequations}
\begin{eqnarray}
    F_{in}=\frac{\omega}{\pi}\int_{0}^{2\pi/\omega}F_{ts}(t)sin(\omega t)\partial t \\
    F_{out}=\frac{\omega}{\pi}\int_{0}^{2\pi/\omega}F_{ts}(t)cos(\omega t)\partial t
\end{eqnarray}
\end{subequations}

\noindent Therefore we can separate Equation~\ref{eq:eom_substituted1} into its orthogonal components and rearrange to find the frequency shift ($\Delta\omega$) and drive amplitude ($F_d$) given any generic tip-sample interaction $F_{ts}$ in the regime where $\omega\approx\omega_o$: 
\begin{subequations}\label{eq:Suppdfdg}
\begin{eqnarray}
    \Delta\omega=\omega-\omega_o = \frac{-\omega_o}{2 kA}\frac{\omega_o}{\pi}\int_{0}^{2\pi/\omega}F_{ts}(t)sin(\omega t)\partial t~~~~ \\
    F_d = \frac{kA}{Q}-\frac{\omega_o}{\pi}\int_{0}^{2\pi/\omega}F_{ts}(t)cos(\omega t)\partial t ~~~~
\end{eqnarray}
\end{subequations}

\noindent Consequently, that component of the tip-sample force that is in phase with the cantilever position shifts the cantilever resonant frequency, and the out-of-phase component of the force contributes to the drive signal. Finally, in the regime where $\omega\approx\omega_o$ the drive signal may be expressed as an energy loss per cycle as\cite{Morita2002}:

\begin{equation}
    E_{ts} = E_o\left[\frac{F_d-F_{do}}{F_d}\right]
\end{equation}

\noindent where $E_o = \frac{\pi k a^2}{Q}$ is the intrinsic damping loss of a high quality factor oscillator\cite{Morita2002} and $F_{do}=\frac{kA}{Q}$ is the offset in Equation~4b. Experimental drive signals are also converted to units of energy loss per cycle given the above expression, where $F_d=A_{exc}$ is the measured drive amplitude Volts and $F_{do}=A_{exco}$ is the drive amplitude in Volts measured in the absence of a tip-sample interaction.

\newpage
\section{III. Substrate} \label{Sec:Substrate}

Reference bias spectroscopy measurements were taken on the SiO\textsubscript{2} substrate underlying the MoSe\textsubscript{2} sample. The frequency shift, as compared to Figure~1b, is parabolic because Si has a comparatively small band gap of $1.1~eV$ and a larger total insulator thickness ($z_{ins}$) due to the SiO\textsubscript{2} overlayer. Section V of the Supplemental Material (Figures~\ref{fig:FitParameters_Phys}a and \ref{fig:FitParameters_AFM}a, respectively) demonstrates these trends toward parabolicity as the band gap decreases and the insulator thickness increases. In both cases, this is due to a reduction in the nonlinearity of the contact potential as a function of bias. (The larger $z_{ins}$ also explains why the magnitude of the SiO\textsubscript{2} frequency shift is much smaller than on MoSe\textsubscript{2}.) The parabolic frequency shift indicates that there is very little charge reorganization occurring within the material. Therefore, the excitation, as compared to Figure~2 on MoSe\textsubscript{2}, is constant on SiO\textsubscript{2} because there no phase offset between the tip-sample position and the tip-sample force.

\begin{figure}[h!]
    \centering
    \includegraphics[width=0.5\linewidth]{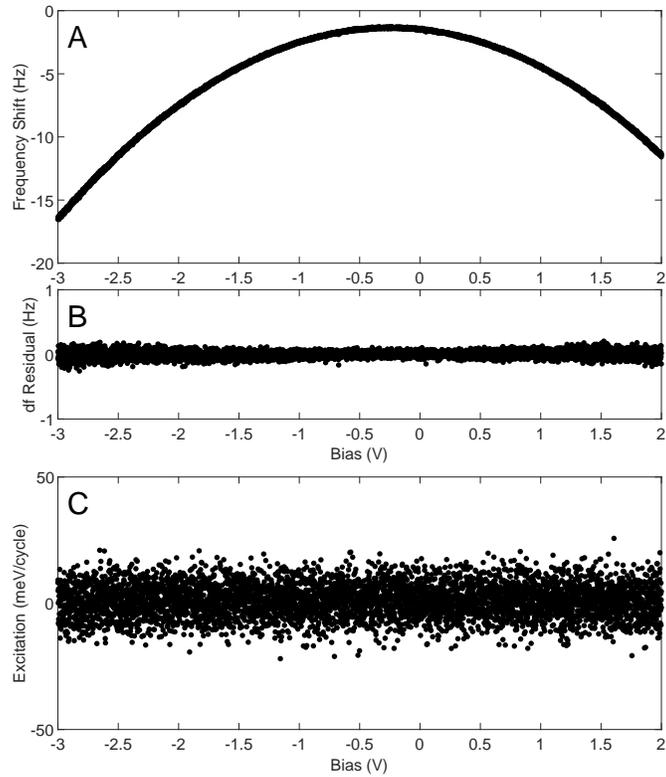}
    \caption{a) Frequency shift measured on the SiO\textsubscript{2} substrate. b) Residual of the frequency shift and a parabolic fit of the form $y=a(x-b)^2+c$ with free fit parameters $a=-1.995$, $b=-0.248$, and $c=-1.359$.  c) Excitation measured simultaneously with the frequency shift shown in (a).}
    \label{fig:Substrate}
\end{figure}

\newpage
\section{IV. Height Dependence} \label{Sec:HeightDependence}
Below are frequency shift and excitation bias spectra at varying zins (i.e. tip lift) above the 
MoSe\textsubscript{2} sample (left) and SiO\textsubscript{2} substrate (right). These experimental results corroborate the
model described in this work (see for comparison Supplementary Material Section V: Fits).

\begin{figure}[h!]
    \centering
    \includegraphics[width=1\linewidth]{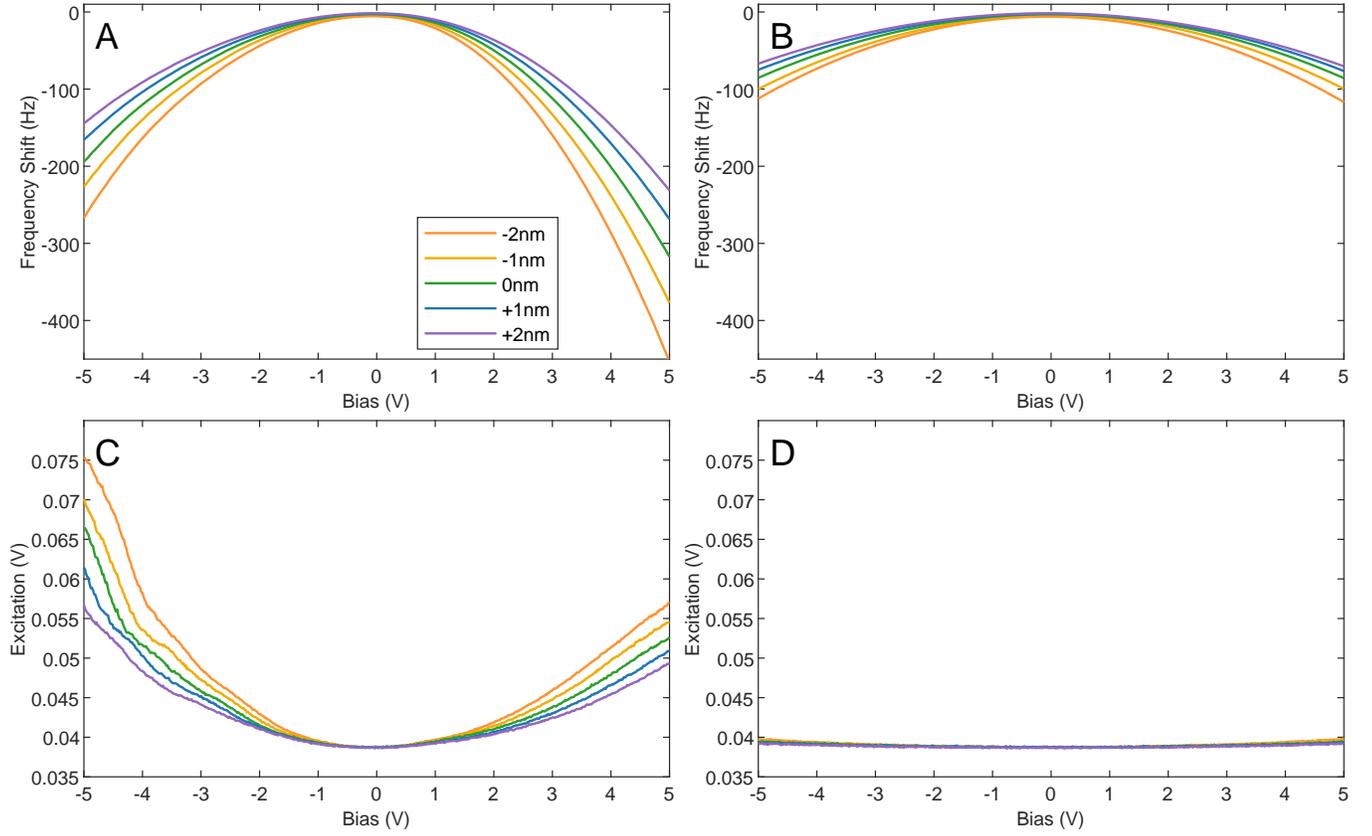}
    \caption{Experimental bias spectra showing the frequency shift (top) and excitation (bottom) at variable tip lift above the MoSe\textsubscript{2} multilayer (a,c) and the SiO\textsubscript{2} substrate (b,d). The scale bars are equal for comparison purposes. Each curve is the average of the forward (negative to positive bias) and backward sweep, above a $-3~Hz$ approach setpoint.}
    \label{fig:HeightDependence}
\end{figure}

\newpage
\section{V. Location Dependence} \label{Sec:LocationDependence}

\begin{figure*}[ht]
    \centering
    \includegraphics[width=\linewidth]{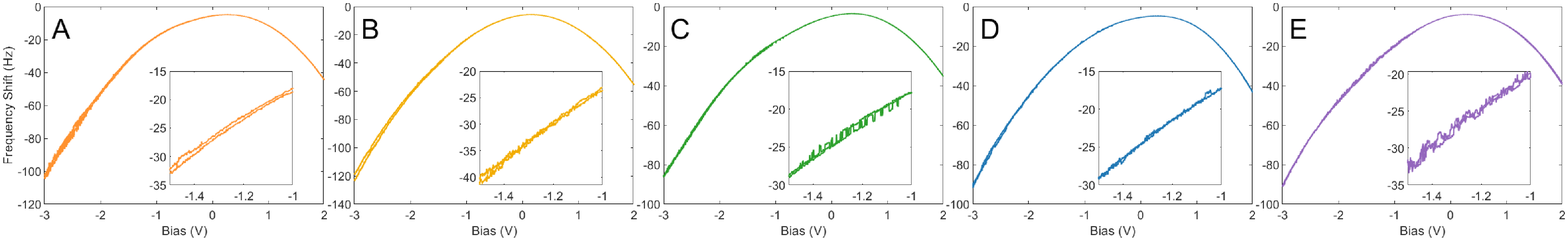}
    \caption{Bias spectra showing the frequency shift at various positions on the MoSt\textsubscript{2} multilayer, spaced between $50$ to $200~nm$ apart. For each (a-e), the approach setpoint was $-3~Hz$, and both the forward (negative to positive bias) and backward sweeps are shown.}
    \label{fig:LocationDependence}
\end{figure*}

Figure~\ref{fig:LocationDependence} demonstrates of the spatial variation in fluctuations that was observed on MoSe2. These
spectra (A-E) were collected at five different locations, ranging between 50nm-200nm apart.
Each image displays two bias sweeps, which confirms that the presence/absence of fluctuations
is robust. The inset of each image shows the region of the parabola between -1.5:-1.0 V, to
highlight that the fluctuations are present in some locations (C, E) but not others (A, B, D)

\begin{figure}[h!]
    \centering
    \includegraphics[trim=0cm 0.3cm 0cm 2cm, clip=true, width=0.8\linewidth]{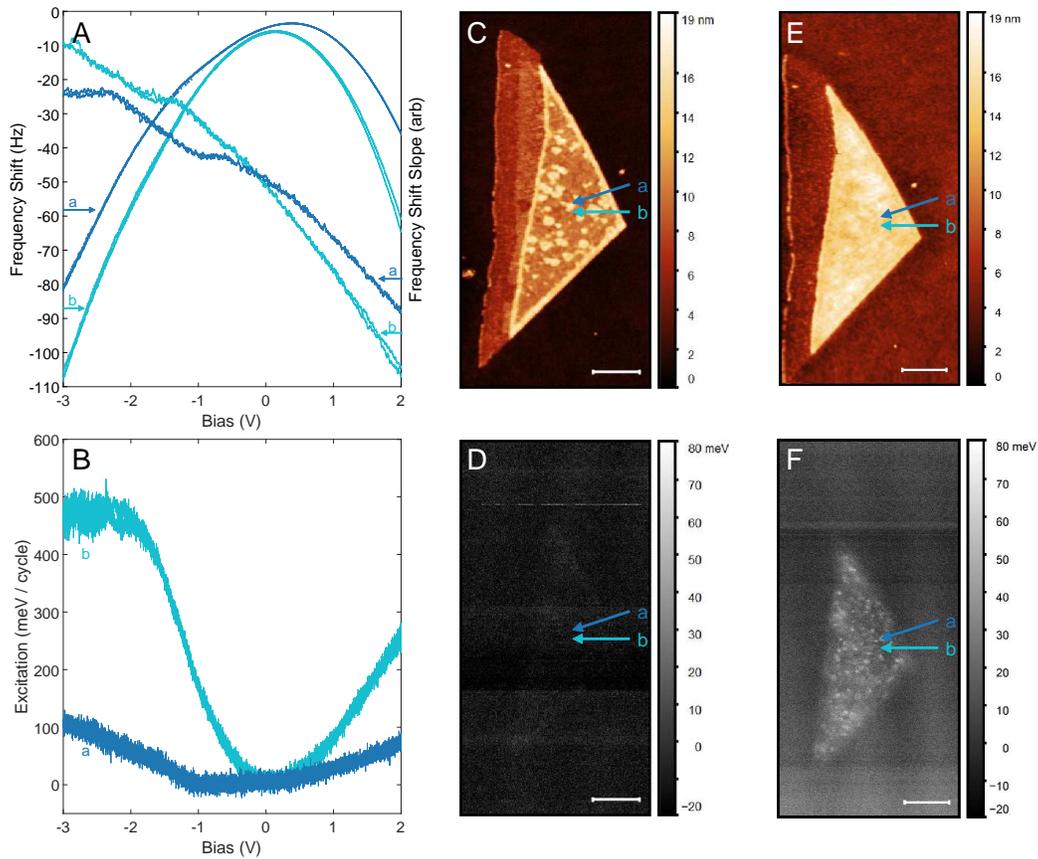}
    \caption{a,b) Frequency shift and excitation bias spectra acquired at two locations labeled `a' and `b', separated by $50~nm$. Both ‘forward’ (negative to positive bias) and ‘backward’ bias curves are shown. c,d) Frequency shift and excitation fm-AFM images at 0V and e,f) -1.5V. The horizontal scale bar corresponds to $500~nm$.}
    \label{fig:SpatialHeterogeneity}
\end{figure}

Figure~\ref{fig:SpatialHeterogeneity} shows the frequency shift and excitation bias spectra (a-b) at two locations, as well as frequency shift and excitation images measured at $0~V$ (c-d) and $-1.5~V$ (e-f). The two positions are separated by approximately $50~nm$. 

\newpage
\section{VI. fm-AFM Signal Source} \label{Sec:SignalSource}

\begin{figure*}[ht]
    \centering
    \includegraphics[width=\textwidth]{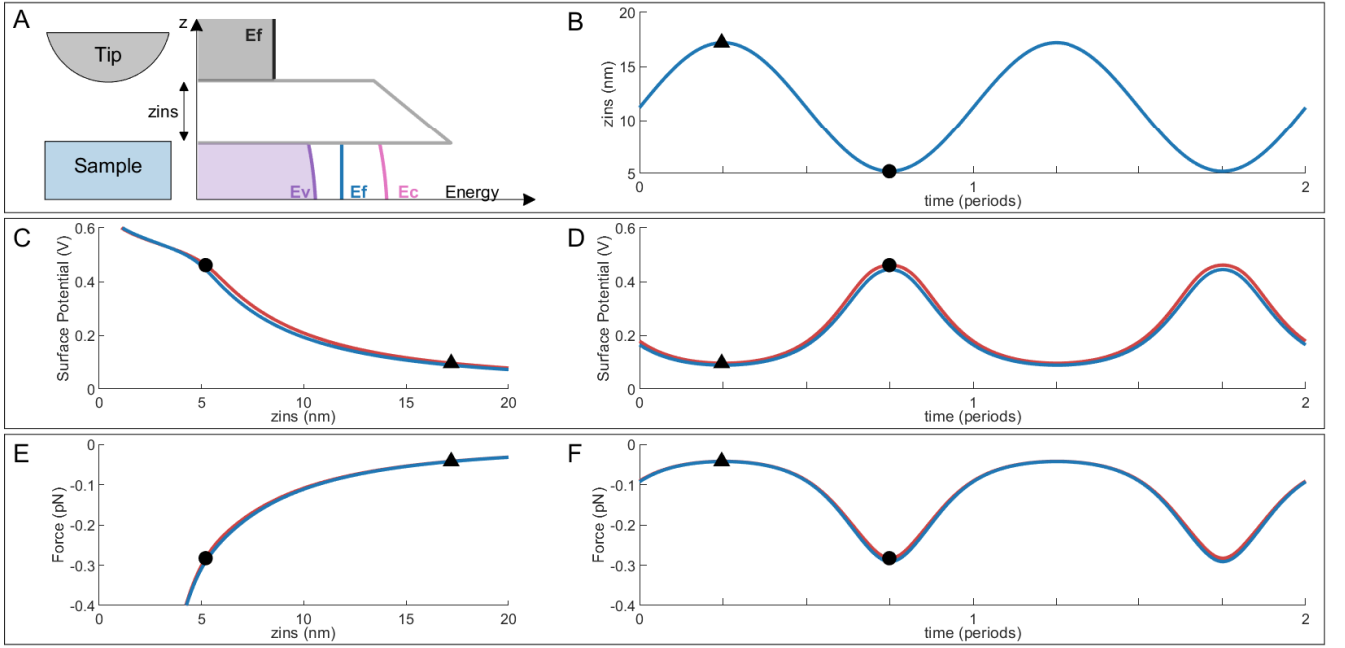}
    \caption{Simulation of the origin of a fm-AFM force measured over a semiconducting sample at a bias of $-1.4~V$ for two acceptor concentrations: $6.3 \mathrm{e}17/cm^3$ (black) and $7.1 \mathrm{e}17/cm^3$ (grey). (The remaining parameters values are the best fit values defined in the main text.) a) Schematic of the fm-AFM setup corresponding to a band diagram of an MIS capacitor oriented along the z-axis. Pink (E\textsubscript{c}) and purple (E\textsubscript{v}) are the sample conduction and valence bands, blue (E\textsubscript{f}) shows the sample Fermi level, dark grey (E\textsubscript{f}) shows metallic tip (gate) Fermi level, and light grey demonstrates the potential drop across the insulator (vacuum). b) Time trace showing that the thickness of the insulator $z_{ins}$ (i.e. tip-sample separation) varies sinusoidally over every oscillation cycle. c,d) The surface potential as a function of $z_{ins}$ and time. e,f) The tip-sample force as a function of $z_{ins}$ and time. Two points, a light green circle and a dark green triangle, are shown for b-f to clarify the relationship between each subfigure. See additional Supplemental Material\cite{SupplementalVideo} for an animated version of this figure.}
    \label{fig:SignalSource}
\end{figure*}

To model the frequency shift and excitation channels measured in a fm-AFM experiment, the time-dependent force (which varies over every oscillation cycle, as tip-sample separation $z_{ins}$ varies) must be known. The time-dependent force is calculated by determining the surface potential at varying $z_{ins}$, according to Equation~\ref{eq:Continuity}, and then calculating the force according to Equation~\ref{eq:Force_MIS}. The $z_{ins}$ and time dependencies of the tip-sample separation, surface potential, and force are shown in Figure~\ref{fig:SignalSource}. An animated version of Figure~\ref{fig:SignalSource} is provided in the additional Supplemental Material\cite{SupplementalVideo}. To calculate the models shown in Figure~\ref{fig:BiasSweeps}a-b, the $V_S(t)$ and $F(t)$ curves shown in Figure~\ref{fig:SignalSource} are calculated at varying bias. For each bias, the frequency shift and excitation are calculating by integrating $F(t)$ according to Equation~\ref{eq:FreqShiftDissipation}.

\newpage
\section{VII. Fits} \label{Sec:Fits}

\begin{figure*}[h!]
    \includegraphics[width=0.95\textwidth]{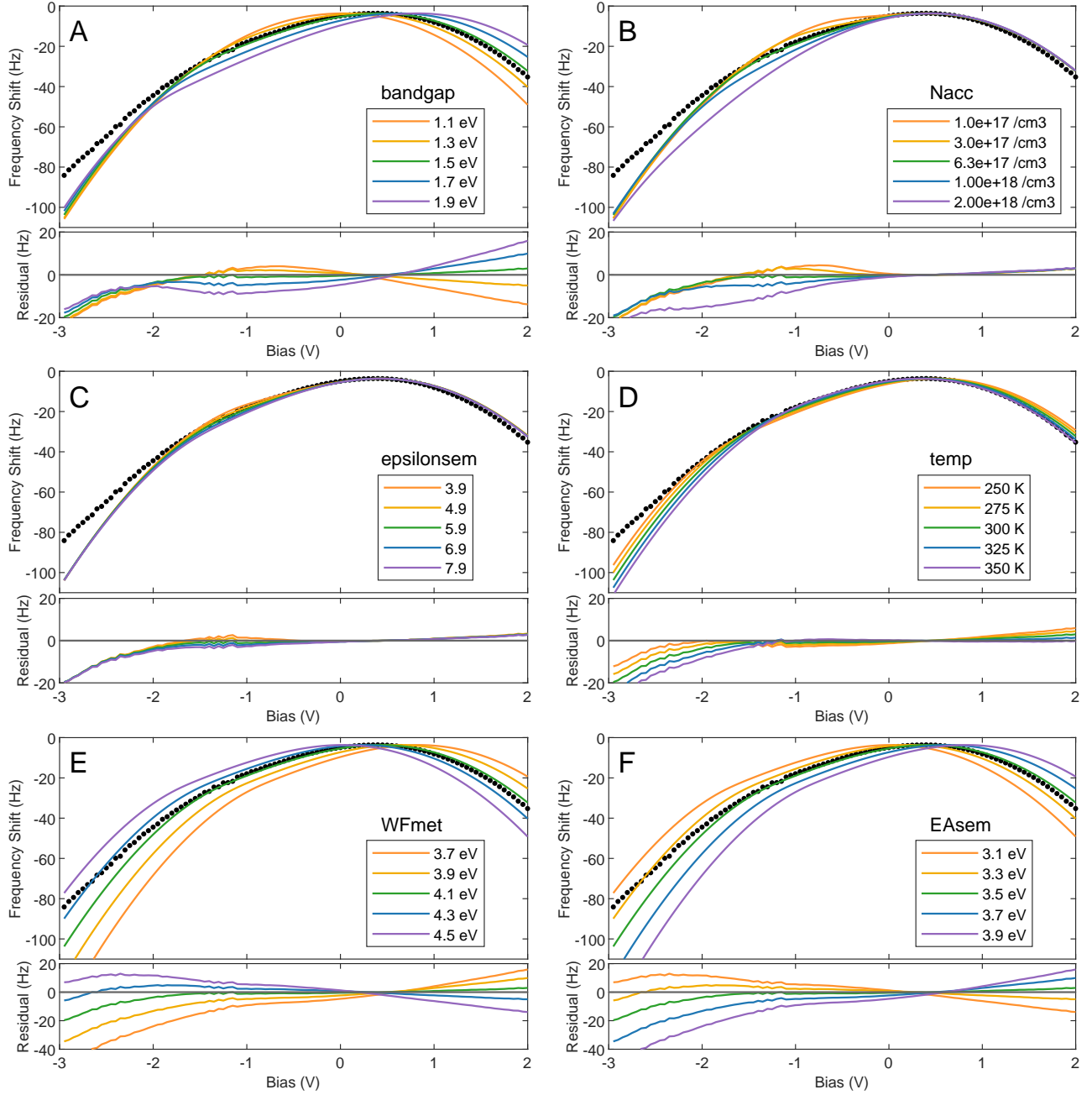}
    \caption{Visualization of the sensitivity of generic MIS system fit parameters: a) band gap $E_g$; b) Acceptor concentration $N_A$; c)  Permittivity $\epsilon$; d) Temperature $T$; e) Metal work function; f) Semiconductor electron affinity. The data from Figure~1 is shown in black. The best fit values are shown in green.} 
    \label{fig:FitParameters_Phys}
\end{figure*}

\begin{figure*}[h!]
    \includegraphics[width=0.95\textwidth]{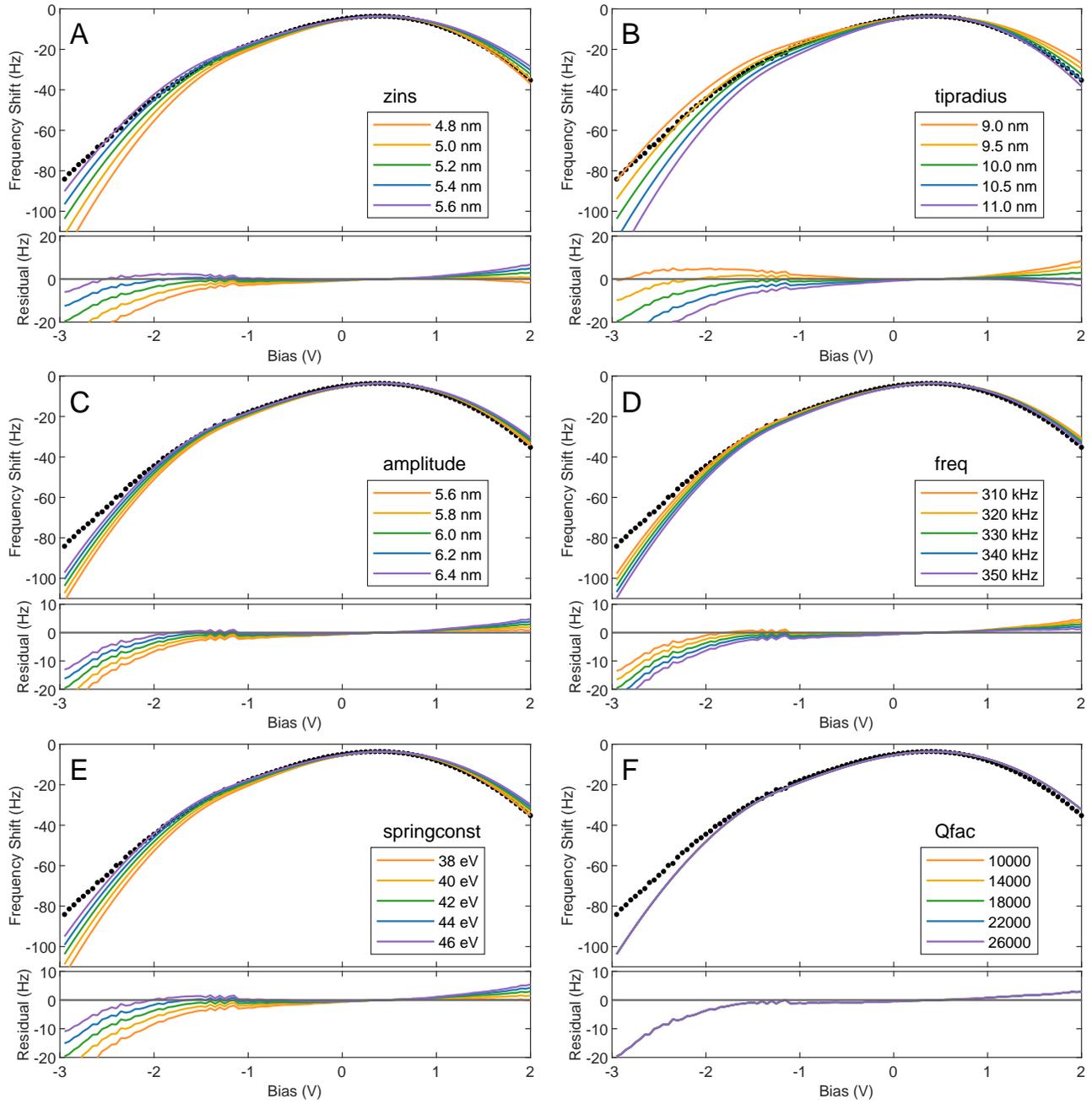}
    \caption{Visualization of the sensitivity of fm-AFM-specific fit parameters (which are all multiplicative prefactors): a) tip-sample separation $z_{ins}$; b) tip radius $r_{tip}$.; c) oscillation amplitude $A$; d) oscillation frequency $\omega$; e) spring constant $k$; f) Q-factor $Q$ (note that Q-factor curves are all perfectly overlapping). The data from Figure~1 is shown in black. The best fit values are shown in green.} 
    \label{fig:FitParameters_AFM}
\end{figure*}

Of the total fourteen experimental parameters in this fit, only three material parameters affect the overall shape of the frequency shift. These are the band gap $E_g$, acceptor concentration $N_A$, and permittivity $\epsilon$. The tip work function and sample electron affinity simply provide lateral shifts (Equation~\ref{eq:Suppcont}) in the frequency shift and leave the shape unchanged. There are six fmAFM-specific parameters (tip-sample separation $z_{ins}$, tip radius $r_{tip}$, oscillation amplitude $A$, oscillation frequency $\omega$, spring constant $k$, and Q-factor $Q$) which all are simply scaling constants (see Equations~\ref{eq:SuppFts} and \ref{eq:Suppdfdg}a). Slight variations around the best fit values for each of these parameters are shown in Figures~\ref{fig:FitParameters_Phys} and \ref{fig:FitParameters_AFM}, to visually demonstrate the sensitivity of the frequency shift curve to each parameter.

\newpage

\newpage